# Optimal Networks from Error Correcting Codes

Ratko V. Tomic
Infinetics Technologies, Inc.
rvtomic@omegapoint.com

## ABSTRACT

To address growth challenges facing large Data Centers and supercomputing clusters a new construction is presented for scalable, high throughput, low latency networks. The resulting networks require 1.5-5 times *fewer switches*, 2-6 times *fewer cables*, have 1.2-2 times *lower latency* and correspondingly lower congestion and packet losses than the best present or proposed networks providing the same number of ports at the same total bisection. These advantage ratios increase with network size.

The key new ingredient is the *exact equivalence* discovered between the problem of *maximizing network bisection* for large classes of practically interesting Cayley graphs and the problem of *maximizing codeword distance* for linear error correcting codes. Resulting translation recipe converts existent optimal error correcting codes into optimal throughput networks.

Ethernet implementation was developed and a prototype built using managed COTS switches. Integrated control plane handles topology, distribution of forwarding tables and fault recovery. Scalable routing uses stretch-free topological addressing. Local load balancing distributes flows at the source over multiple, non-minimal, edge disjoint paths. Path selection does not use tunneling or overlays but embeds path selectors in the topological addresses resulting in wire-speed forwarding and allowing for cut-through switching where available.

## Categories and Subject Descriptors

C.2.1 [**Computer Communication Networks**]: Network Architecture and Design – *network topology, packet switching networks*; E.4 [**Coding and Information Theory**]: Error control codes; G.2.2 [**Discrete Mathematics**]: Graph Theory – *graph algorithms, network problems*.

## 1. INTRODUCTION

Rapid growth of Data Centers (DC) along with rise in virtualization, cloud and Big Data services, all boosting intra-DC traffic, has stressed capabilities of conventional 'three tier' DC architecture sparking a flurry of proposals for new DC designs [1]- [9].

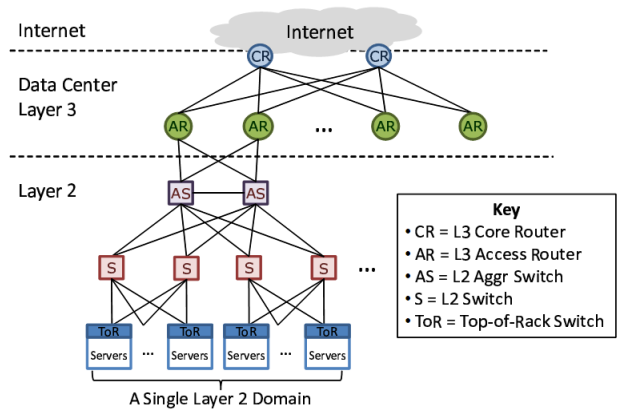

**Figure 1-1: Conventional Data Center**, [3]

At the root of conventional DC problems is non-scalable Layer 2 (L2) with fragmented control plane using flood based coordination (ARP) and forwarding. That approach constrains L2 to a loopless topology, tree, which limits the size of L2 domains, creating bottlenecks and requiring expensive high radix switches at the root of the tree. To grow a DC beyond few thousand servers, multiple L2 domains are connected as subnets into a Layer 3 network via large, expensive routers, increasing oversubscriptions to as high as 200, hampering agility, mobility and resulting in labor intensive network management.

The solution presented, ***Flexible Radix Switch***™ (FRS)[1], addresses the root DC problems, the fragmented, non-scalable control plane and tree topology[2]. The name FRS reflects the high degree of integration of network resources, from fabric and wiring aggregation via a novel, mathematically optimal[3] ***Long Hop***™ topology (LH), through integrated control and management planes with factory like division of labor and maximum pooling of common functions and resources. The resulting network appears functionally as a single high throughput, low latency switch with flexible radix, capable of scaling the single flat L2 domain to practically any size Data Center. The net effect on DC economy of FRS built from existent ToR switches (or managed COTS switches) connected via Long Hop topology is shown below – while lowering the oversubscription 20X, the aggregation layers of switches and routers are made unnecessary, along with the associated cabling and power.

---

[1] Flexible Radix Switch and Long Hop are trademarks of Infinetics Technologies, Inc.
[2] In common with [2], [4], FRS was inspired by HPC systems.
[3] Within a large class of symmetrical networks (cf. sec. 3).

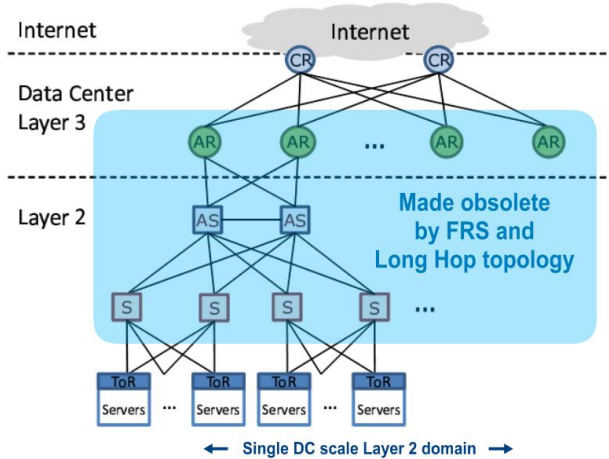

**Figure 1-2: FRS Economy**

The integrated control and management planes of FRS utilize similar ideas and techniques as those used by other proposals [2] to [9], hence most of the paper is focused on the key new advance, the Long Hop topology.

## 2. MATHEMATICAL TOOLS

Since the methods used cross several disciplines not usually brought together, this section introduces terms and results needed in a harmonized notation.

### 2.1 Terms and Notation

- $\mathbb{V}_n$ – $n$-dimensional vector space over implicit field $\mathbf{F}_q$
- $\mathbb{S}(k,n,q)$ – $k$-dimensional subspace of $\mathbb{V}_n$ (linear span) over field $\mathbf{F}_q$. Also: $\mathbb{S}(k,n)$ for implicit $\mathbf{F}_q$.
- $\langle \mathbf{X}| \equiv (x_1\ x_2 \cdots x_n)$ – row vector (Dirac notation [10])
- $|\mathbf{Y}\rangle \equiv (y_1\ y_2 \cdots y_n)^T$ – column vector ($^T$ is 'transposed')
- $\langle \mathbf{X}|\mathbf{Y}\rangle \equiv \sum_{i=1}^{n} x_i\ y_i$ – scalar product of vectors $\mathbf{X}$ and $\mathbf{Y}$
- $\mathbf{A} \equiv |\mathbf{Y}\rangle\langle\mathbf{X}|$ – matrix $\mathbf{A}$ with elements $\mathbf{A}_{i,j} = y_i x_j$
- $\langle e_i| \equiv (0_1 0_2 \cdots 0_{i-1}\ 1_i\ 0_{i+1} \cdots 0_n)$ – std. basis vector
- $\mathbf{I}_n$ – $n \times n$ identity matrix
- $a\ \%\ b$ – integer $a$ modulo integer $b$, same as: $a$ mod $b$
- $\tilde{a}$ – bitwise complement of bit string $a$
- $a\ \&\ b$ – bitwise AND of $a$ and $b$
- $a\ |\ b$ – bitwise OR of $a$ and $b$
- $a\ \hat{}\ b$ – bitwise XOR of $a$ and $b$
- $[E]$ – Iverson bracket = 1 (or 0) if $E$ true (or false)
- $\delta_{i,j}$ – Kronecker delta, same as $[i=j]$
- $\mathbf{A} \otimes \mathbf{B}$ – Kronecker product of matrices $\mathbf{A}$ and $\mathbf{B}$
- $\mathbf{A} \oplus \mathbf{B}$ – Direct sum of matrices (of vector spaces)
- iff – "if and only if"

**Binary expansion** of a $d$-bit integer $X = \sum_{\mu=0}^{d-1} x_\mu\ 2^\mu \equiv x_{d-1} \ldots x_1 x_0$ (bit string form)

**Parity** of a $d$-bit integer $\mathbf{X} = x_{d-1} \ldots x_1 x_0$ is defined as:
$\mathbb{P}(\mathbf{X}) \equiv (x_0 + x_1 + \ldots + x_{d-1})\ \text{mod}\ 2 = x_0\ \hat{}\ x_1\ \hat{}\ \ldots\ \hat{}\ x_{d-1}$

**Hamming weight** $\Delta\mathbf{X}$ of $n$-tuple $\mathbf{X} \equiv x_1\ x_2 \ldots x_n$ is the number of non-zero symbols in $\mathbf{X}$. **Hamming distance** $\Delta(\mathbf{X},\mathbf{Y})$ between $n$-tuples $\mathbf{X}$ and $\mathbf{Y}$ is the number of positions $i$ where $x_i \neq y_i$; hence $\Delta\mathbf{X} = \Delta(\mathbf{X},\mathbf{0})$.

**Cyclic group** $\mathbf{Z}_n$: set of integers $\{0,1,\ldots n-1\}$ with integer addition modulo $n$ as the group operation.

**Product group** $\mathbf{Z}_q^d \equiv \mathbf{Z}_q \times \mathbf{Z}_q \times \cdots \times \mathbf{Z}_q$ ($d \times$): extension of $\mathbf{Z}_q$ into a $d$-tuple. Group $\mathbf{Z}_2^d$ is a group of $d$-bit strings with bitwise XOR as the group operation.

### 2.2 Walsh Functions

Hadamard matrix $\mathbf{H}_n$ is a symmetric matrix defined for power of two sizes $n=2^d$ by the recursion (cf. [11]):

$$\mathbf{H}_2 = \begin{pmatrix} 1 & 1 \\ 1 & -1 \end{pmatrix},\ \mathbf{H}_{2n} = \begin{pmatrix} \mathbf{H}_n & \mathbf{H}_n \\ \mathbf{H}_n & -\mathbf{H}_n \end{pmatrix} \quad (2.1)$$

Walsh functions $\mathbf{U}_r(x)$ are defined for $r, x \in [0, n)$ via the elements of Hadamard matrix $\mathbf{H}_n$ as follows:

$$\mathbf{U}_r(x) \equiv (\mathbf{H}_n)_{r,x} \quad (2.2)$$

*Walsh vector* $\mathbf{U}_r \equiv \langle \mathbf{U}_r| \equiv (\mathbf{U}_r(0)\ \mathbf{U}_r(1) \ldots \mathbf{U}_r(n-1))$ is thus the $r$-th row of $\mathbf{H}_n$. Some properties of $\mathbf{U}_r(x)$ needed later are:

*Orthogonality*: $\langle \mathbf{U}_r|\mathbf{U}_s\rangle = n \cdot \delta_{r,s} = \begin{cases} n & \text{for } r = s \\ 0 & \text{for } r \neq s \end{cases}$ (2.3)

*Symmetry*: $\mathbf{U}_r(x) = \mathbf{U}_x(r)$ (2.4)

$$\mathbf{U}_r(x) = (-1)^{\sum_{\mu=0}^{d-1} r_\mu x_\mu} = (-1)^{\mathbb{P}(r\&x)} \quad (2.5)$$

$$\langle \mathbf{U}_0| = (1\ 1\ \ldots 1) \equiv \langle 1| \quad (2.6)$$

$$\sum_{x=0}^{n-1} \mathbf{U}_r(x) = 0,\ r = 1..n-1 \quad (2.7)$$

Eq. (2.7) shows that each vector $\langle \mathbf{U}_r|$ for $r>0$ has equal numbers of +1 and -1 elements. For implementations in software or hardware a binary form $\mathbf{W}_r(x)$ of $\mathbf{U}_r(x)$, which replaces $1 \to 0$ and $-1 \to 1$, is often more useful. Algebraic values $a \equiv \mathbf{U}_r(x)$ are related to binary values $b \equiv \mathbf{W}_r(x)$ as:

$$b \equiv \frac{1-a}{2},\ a = 1 - 2b \quad (2.8)$$

The function values of $\mathbf{W}_r(x)$ from eq. (2.5) are:

$$\mathbf{W}_r(x) = \mathbb{P}\left(\sum_{\mu=0}^{d-1} r_\mu x_\mu\right) = \mathbb{P}(r\&x) \quad (2.9)$$

Eq. (2.9) and properties of binary operators imply:

$$\mathbf{W}_r(x)\ \hat{}\ \mathbf{W}_s(x) = \mathbf{W}_{r\hat{}s}(x) \quad (2.10)$$

### 2.3 Matrices and Eigenvectors

For matrix $\mathbf{A}$, *eigenvector* $|X\rangle$ is any solution of equation:

$$\mathbf{A}|X\rangle = \lambda\ |X\rangle \quad (2.20)$$

where $\lambda$ is a scalar value called *eigenvalue* of $\mathbf{A}$ for $|X\rangle$.

(**M**₁) All symmetric real-valued $n \times n$ matrices **A** have $n$ real eigenvalues and $n$ orthogonal eigenvectors which form a basis (*eigenbasis*) in $\mathbb{V}_n$.

(**M**₂) A set of $m$ real, symmetric, pairwise commuting matrices $\mathcal{F}_m \equiv \{ \mathbf{S}_k : \mathbf{S}_k \mathbf{S}_j = \mathbf{S}_j \mathbf{S}_k \text{ for } j,k = 1..m \}$ is called *commuting family*. Any commuting family $\mathcal{F}_m$ has an orthonormal set of $n$ vectors (eigenbasis) $\{|v_i\rangle\}$ which are simultaneously eigenvectors of all $\mathbf{S}_k \in \mathcal{F}_m$ ( [12] p. 52).

(**M**₃) Labeling $n$ real eigenvalues of a symmetric matrix **A** as: $\lambda_{min} \equiv \lambda_1 \leq \lambda_2 \leq \cdots \leq \lambda_n \equiv \lambda_{max}$, then the following equalities hold (*Rayleigh-Ritz* theorem, cf. [12] p. 176):

$$\lambda_{min} \equiv \lambda_1 = \min_{X \in \mathbb{V}_n} \left\{ \frac{\langle X|\mathbf{A}|X\rangle}{\langle X|X\rangle}, X \neq 0 \right\} \quad (2.21)$$

$$\lambda_{max} \equiv \lambda_1 = \max_{X \in \mathbb{V}_n} \left\{ \frac{\langle X|\mathbf{A}|X\rangle}{\langle X|X\rangle}, X \neq 0 \right\} \quad (2.22)$$

In words – the min/max values of the '*Rayleigh quotient*' $\mathbf{M}_V \equiv \langle X|\mathbf{A}|X\rangle / \langle X|\mathbf{A}|X\rangle$ are solved by the eigenvector $\mathbf{X}_0$ of **A** corresponding to $\lambda_{min}$ or $\lambda_{max}$ eigenvalues of **A**.

(**M**₄) Decomposing[4] space $\mathbb{V}_n = \mathbb{V}_1(\mathbf{X}_0) \oplus \mathbb{V}_{n-1}$ from (**M**₃) and applying (**M**₃) to $\mathbb{V}_{n-1}$ solves the min/max problem for subspace $\mathbb{V}_{n-1}$ with the next eigenvector, corresponding to $\lambda_2$ or $\lambda_{n-1}$ (*Courant-Fisher* theorem, cf. [12] p. 179).

## 2.4 Cayley Graphs

A graph $\Gamma(\mathsf{V},\mathsf{E})$ is an object with $n$ *vertices* (nodes) $\mathsf{V} = \{v_1, v_2, \ldots v_n\}$ and $c$ *edges* (links) $\mathsf{E} = \{\varepsilon_1, \varepsilon_2, \ldots \varepsilon_c\}$, where each edge $\varepsilon$ is (connects) a pair of vertices. We will consider only *undirected* graphs (bidirectional links). *Node degree* (topological radix), denoted as $m$, is number of links connected to a node.

*Adjacency matrix* **A** of a graph is $n \times n$ matrix with elements $A_{i,j} = 1$ if $v_i$ and $v_j$ are connected, 0 otherwise. For undirected graphs **A** is always a *symmetric* matrix. For graphs of interest here with fixed $m$ for all nodes (*regular graphs*), each row and column of **A** has $m$ ones, hence:

$$\sum_{j=1}^{n} \mathbf{A}_{i,j} = \sum_{i=1}^{n} \mathbf{A}_{i,j} = m \quad (2.25)$$

$$\sum_{i,j=1}^{n} \mathbf{A}_{i,j} = n \cdot m \quad (2.26)$$

Of particular interest for networking are *Cayley graphs* (**CG**) due to their vertex symmetry (network looks the same from each node which reduces computations), simple, regular construction and routing (often self-routing), good scaling properties, low latencies and high bisections for given node degree [13], [14]. Some better known Cayley graphs are hypercube, folded cube, cube connected cycles, hyper-torus, flattened butterfly, HyperX [15], star graph, complete graph, transposition graphs, etc.

---

[4] $\mathbb{V}_1(\mathbf{X}_0)$ is 1-dimensional space spanned by vector $\mathbf{X}_0$ and $\mathbb{V}_{n-1}$ is $n$-1 dim. space spanned by the $n$-1 remaining eigenvectors.

*Cayley graph* $Cay(\mathbf{G}_n, \mathbf{S}_m)$ is defined via a *group* $\mathbf{G}_n$ of $n$ elements $\{ g_1 \equiv \mathbf{I}, g_2, \ldots g_n \}$ and its proper subset $\mathbf{S}_m = \{ h_1, h_2, \ldots h_m \}$ called *generator set* satisfying (cf. [16] ch. 5):

**CG**₁) for any $h \in \mathbf{S}_m \Rightarrow h^{-1} \in \mathbf{S}_m$ (bidirectionality)
**CG**₂) $\mathbf{S}_m$ does *not* contain identity **I** (no self-loops)

**CG construction:** $Cay(\mathbf{G}_n, \mathbf{S}_m)$ has $\mathsf{V} \equiv \{ g_1, g_2, \ldots g_n \}$ and the edge set is $\mathsf{E} \equiv \{ (g_i, g_i \cdot h_s), \forall i, s \}$. In words, each node $g_i$ is connected to $m$ nodes $\{ g_i \cdot h_s, s = 1..m \}$.

Generating elements $h_s$ are called here *hops* since for identity element $g_1 \equiv \mathbf{I}$ (*root node*) their group action is precisely the single hop transition from the root node $g_1$ to its 1-hop neighbors $h_1, h_2, \ldots h_m \in \mathsf{V}$.

Construction of folded 3-cube $FQ_3 = Cay(\mathbf{Z}_2^3, \mathbf{S}_4)$ is shown in Figure 2-1. The group is $n = 8$ element group $\mathbf{Z}_2^3$ and generator set is $\mathbf{S}_4 = \{001, 010, 100, 111\}$ (labels are in binary). Arrows on the links indicate group action (XORs node labels with generators) on vertex $v_1 = 000$ (identity, root). The requirement $CG_1$ follows from the involution of the XOR operation: $x \wedge x = 0$, i.e. each hop $h_i$ is its own inverse (since identity element of $\mathbf{Z}_2^d$ is $\mathbf{I} = 0$).

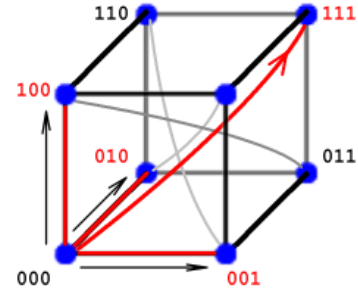

**Figure 2-1: Folded 3-cube**

## 2.5 Error Correcting Codes

Error correcting codes (ECC) are techniques for adding redundancy to messages in order to detect or correct errors in the decoding phase. Of interest here are the *linear EC codes*, which are the most developed and in practice the most important type of ECC [17], [18].

*Message* **X** is a sequence of $k$ symbols $x_1, x_2, \ldots, x_k$ from alphabet $\mathcal{A}$ of size $q \geq 2$ i.e. $x_i$ can be taken to be integers with values in interval $[0,q)$. EC code for **X** is a *codeword* **Y** which is a sequence $y_1, y_2, \ldots, y_n$ of $n > k$ symbols from $\mathcal{A}$. Encoding procedure translates all messages from some set $\{\mathbf{X}\}$ into codewords from some set $\{\mathbf{Y}\}$. For *block codes* the sizes of the sets $\{\mathbf{X}\}$ and $\{\mathbf{Y}\}$ are $q^k$ i.e. **X** is an arbitrary $k$-tuple in alphabet $\mathcal{A}$. The excess $n-k > 0$ symbols in **Y** are called *coding redundancy* or "check bits" that support detection or correction of errors during decoding of **Y** into **X**.

For ECC algorithmic purposes, alphabet $\mathcal{A}$ is augmented with additional mathematical structure, beyond that of a set. Common augmentation is to view symbols $x_i$ and $y_i$ as elements of *Galois field* **GF**($q$) where $q \equiv p^m$ for a prime

$p$ and an integer $\mathbf{m} \geq 1$[5]. Codewords $\mathbf{Y}$ are then a subset of all $n$-tuples $\mathbf{F}_q^n$ over the field $\mathbf{GF}(q)$. The addition of $n$-tuples $\mathbf{F}_q^n$ and their multiplication with $\mathbf{GF}(q)$ elements is done component-wise i.e. $\mathbf{F}_q^n$ is a finite $n$-dimensional vector space $\mathbb{V}_n \equiv \mathbf{F}_q^n$ over finite field $\mathbf{GF}(q)$.

*Linear EC codes* are a special case of the above $n$-tuple $\mathbf{F}_q^n$ structure of codewords, in which the set $\{\mathbf{Y}\}$ of all codewords is a $k$-dimensional vector *subspace* (or **linear span**) $\mathbb{S}(k,n,q)$ of $\mathbb{V}_n$. Hence, if $\mathbf{Y}_1$ and $\mathbf{Y}_2$ are codewords, then $\mathbf{Y}_3 = \mathbf{Y}_1 + \mathbf{Y}_2$ is also a codeword. The number of distinct codewords $\mathbf{Y}$ in $\mathbb{S}(k,n,q)$ is $|\mathbb{S}(k,n,q)| = q^k$. Linear code is denoted by convention as $[n,k]_q$ or just as $[n,k]$.

A code $[n,k]_q$ is uniquely specified by its $\mathbb{S}(k,n,q)$ which can be constructed from a basis of $k$ linearly independent $n$-dimensional vectors $\langle g_i| = (g_{i,1}\ g_{i,2} \dots g_{i,n})$, $i = 1..k$. This basis defines the $k \times n$ *generator matrix* $\mathbf{G}$ of the $[n,k]_q$ code as follows (cf. [18] p. 84):

$$\mathbf{G} \equiv \sum_{i=1}^{k} |e_i\rangle\langle g_i| = \begin{pmatrix} \langle g_1| \\ \dots \\ \langle g_k| \end{pmatrix} = \begin{pmatrix} g_{1,1} & \dots & g_{1,n} \\ \dots & \dots & \dots \\ g_{k,1} & \dots & g_{k,n} \end{pmatrix} \quad (2.30)$$

i.e. the $k$ row vectors $\langle g_i|$ are the $k$ rows of matrix $\mathbf{G}$. Encoding of some message $\mathbf{X} \equiv \langle \mathbf{X}| \equiv (x_1\ x_2 \dots x_k)$ into codeword $\mathbf{Y} \equiv \langle \mathbf{Y}| \equiv (y_1\ y_2 \dots y_n)$ is defined via:

$$\langle \mathbf{Y}| \equiv \langle \mathbf{X}|\mathbf{G} = \sum_{i=1}^{k} \langle \mathbf{X}|e_i\rangle\langle g_i| = \sum_{i=1}^{k} x_i \langle g_i| \quad (2.31)$$

The most developed and the most useful are binary $[n,k]$ codes using $\mathbf{GF}(2)^n$ as the codeword space to encode $k$-bit binary strings into $n$-bit codewords. Vector additions in $\mathbf{GF}(2)^n$ are XORs of $n$-bit strings. Example: eq. (2.32) shows the Hamming [7,4] code encoding a 4-bit message $\mathbf{X} = 0011$ into a 7-bit codeword $\mathbf{Y}(\mathbf{X}) = 0100011$:

$$(0011)\begin{pmatrix} 1 & 1 & 0 & \mathbf{1} & 0 & 0 & 0 \\ 0 & 1 & 1 & 0 & \mathbf{1} & 0 & 0 \\ 1 & 1 & 1 & 0 & 0 & \mathbf{1} & 0 \\ 1 & 0 & 1 & 0 & 0 & 0 & \mathbf{1} \end{pmatrix} = (0100011) \quad (2.32)$$

As prescribed by eq. (2.31), the positions of 1s in X indicate the positions of rows of matrix $\mathbf{G}$ (last 2 rows) which are XOR-ed to produce the 7-bit codeword $\mathbf{Y}(\mathbf{X})$.

Choice of vectors $\langle g_i|$ used to construct $\mathbf{G}$ depends on type of errors that the $[n,k]$ code is supposed to detect or correct. For the most common assumption in ECC theory, the *independent random errors* for symbols of codeword $\mathbf{Y}$, the best choice of $\langle g_i|$ are those that maximize the minimum Hamming distance $\Delta(\mathbf{Y}_1, \mathbf{Y}_2)$ among all pairs of distinct codewords $\mathbf{Y}_1 \neq \mathbf{Y}_2$. Defining minimum codeword distance $\Delta$ via:

$$\Delta \equiv \min\{\Delta(\mathbf{Y}_1, \mathbf{Y}_2) \mid \forall\, \mathbf{Y}_1, \mathbf{Y}_2 \in \mathbb{S}(k,n,q)\} \quad (2.33)$$

the $[n,k]$ code is often denoted as $[n,k,\Delta]$. The optimum choice for vectors $\langle g_i|$ maximizes $\Delta$ for given $n$, $k$ and $q$.

Tables of optimum and near optimum $[n,k,\Delta]_q$ codes have been computed over decades for wide range of parameters $n$, $k$ and $q$ (e.g. see web repositories [19], [20]).

Quantity related to $\Delta$ of importance for our construction is the *minimum non-zero codeword weight* $\mathbf{w}_{min}$ defined via Hamming weight $\Delta\mathbf{Y}$ as follows:

$$\mathbf{w}_{min} \equiv \min_{\mathbf{Y} \neq 0}\{\Delta\mathbf{Y} : \mathbf{Y} \in \mathbb{S}(k,n,q)\} \quad (2.34)$$

The property of $\mathbf{w}_{min}$ (cf. [18] p. 83) of interest here is that for any linear code $[n,k,\Delta]_q$:

$$\mathbf{w}_{min} = \Delta \quad (2.35)$$

Applying test (2.34) to the example (2.32) using set of 15 non-zero messages $\mathbf{X}$: $\{0001,..\ 1111\}$ to generate 15 codewords $\mathbf{Y}$ for (2.34), yields $\Delta = \mathbf{w}_{min} = 3$. This distance $\Delta = 3$ implies that Hamming [7,4,3] code can detect any 2-bit error and correct any 1-bit error.

($\mathbf{EC}_1$) Eq. (2.35) implies that the *construction of optimal $[n,k,\Delta]_q$ codes* (codes maximizing $\Delta$) is a problem of finding $k$-dimensional subspace $\mathbb{S}(k,n,q)$ of $n$-dimensional vector space $\mathbb{V}_n$ which maximizes $\mathbf{w}_{min}$ of the $\mathbb{S}(k,n,q)$:

$$\Delta_{opt} = \max_{\mathbb{S} \subset \mathbb{V}_n}\left\{\min_{\mathbf{Y} \neq 0}\{\Delta\mathbf{Y} : \mathbf{Y} \in \mathbb{S}(k,n)\}\right\} \quad (2.36)$$

($\mathbf{EC}_2$) Since any set of $k$ linearly independent vectors $\langle g_i|$ (basis) from $\mathbb{S}(k,n,q)$ generates (spans) the same space $\mathbb{S}(k,n,q)$ of $q^k$ vectors $\mathbf{Y}$, $\mathbf{w}_{min}$ and $\Delta$ are *independent of the choice of the basis* $\{\langle g_i| : i = 1..k\}$. Namely by virtue of uniqueness of expansion of all $q^k$ vectors $\mathbf{Y} \in \mathbb{S}(k,n,q)$ in any basis of $\mathbb{S}(k,n,q)$ and pigeonhole principle, the change of basis merely permutes the mapping $\mathbf{X} \to \mathbf{Y}$, retaining exactly the same set of $q^k$ vectors of $\mathbb{S}(k,n,q)$ and all their properties such as $\Delta$ and $\mathbf{w}_{min}$.

Conclusions ($\mathbf{EC}_1$) and ($\mathbf{EC}_2$) are the key results of ECC theory needed for our construction of optimal networks.

## 3. TOPOLOGY OPTIMIZATION

Networks considered have $N$ nodes (switches) of uniform radix $R$ and *uniform* number of *topological* ports per switch $m$ (node degree). Hence the number of *free* (server) *ports* per switch is uniform value $p = R - m$. The total number of free ports in the network is thus $P = p \cdot N$.

Two principal measures of network performance are *bandwidth* and *latency* [21]. We will focus on bandwidth optimization[6]. Common metric for evaluating bandwidth is *bisection* which is defined as follows, [22]:

Vertex set $\mathsf{V}$ is partitioned into two equal disjoint sets (*equipartition*) $\mathbf{S}_1$ and $\mathbf{S}_2$ with $N_1 = N_2 = N/2$ nodes[7]. A *cut*

---

[5] Condition $q = p^m$ is a necessary condition in order to augment a bare set $\mathcal{A}$ into a finite field $\mathbf{F}_q$ (cf. [17] p. 200).

[6] The resulting very low latency (in hops) was an unexpected side-effect of optimizing topology for bisection.

[7] For brevity we restrict $N$ to even values. Total number of distinct equipartitions is then $|\mathbf{E}| = \frac{1}{2}\binom{N}{N/2}$.

**C(X)** for some partition **X** is the number of links[8] crossing between the sets $S_1$ and $S_2$. Bisection **B** is the *minimum cut* **C(X)** in the *set* **E** *of all equipartitions* **X**.

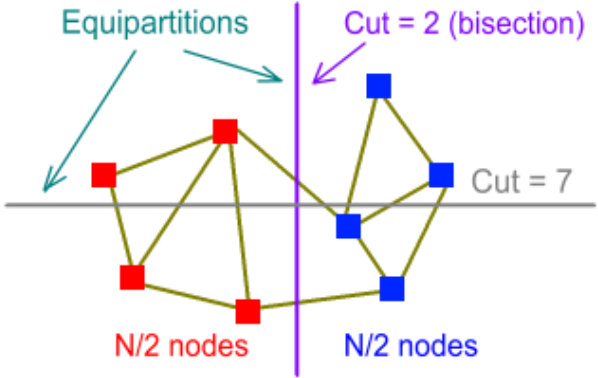

**Figure 3-1: Definition of bisection**

Optimization of network bandwidth given via bisection **B** can be broken into two subproblems:

**P₁**) Find algorithm for **B** for some class of topologies
**P₂**) Maximize this **B** by changing links between nodes[9]

Both problems are intractable for general graphs (NP-complete, [23]) and approximate algorithms for **B** are not simultaneously accurate and scalable enough to serve as a tool for the **P₂**. The best in this class are "entangled networks" computed via simulated annealing in [24], [25]. While achieving a good performance, they could be computed in this manner only to $N \sim 2000$ nodes (with solution quality degrading with size).   ). The Jellyfish topology [26] is unoptimized variant of entagled networks, providing arbitrary sizes but at lower performance. Further, the very high irregularity of such graphs makes them impractical for forwarding, routing, load balancing, parallel algorithm decomposition and physical wiring.

Our approach is to narrow the field to vertex symmetrical graphs already interesting as network topologies, such as Cayley graphs [13], [14], generalize them and solve exactly and efficiently **P₁** and **P₂** for network sizes of interest in the near future ($N < 10^7$ switches). The graphs for which optimal solutions were found include maximum generalizations of hypercube and hyper-torus that retain the Cayley graph symmetry of the original networks.

### 3.1 Computing Bisection

We encode equipartitions of V as vectors $\mathbf{X} \equiv (x_0 \, x_1 \ldots x_{N-1})$ in $\mathbb{V}_N$, where $x_i = \pm 1$ indicates whether node[10] $i$ is in $S_1$ or in $S_2$ half. Hence $x_i \cdot x_j$ is +1 (or -1) if nodes $i$ and $j$ are in the same (or different) halves of V. Since the adjacency matrix element $\mathbf{A}_{i,j}$ is 1 if nodes $i$ and $j$ are connected, 0 otherwise, the expression $\mathbf{C}_{i,j} \equiv \mathbf{A}_{i,j} \cdot (1 - x_i \cdot x_j)/2$ has value $\mathbf{C}_{i,j}=1$ iff nodes $i$ and $j$ are connected ($\mathbf{A}_{i,j}=1$) *and* are in different halves ($x_i \, x_j = -1$), otherwise $\mathbf{C}_{i,j}=0$ (since $\mathbf{A}_{i,j}=0$ or $x_i \cdot x_j=+1$). Hence $\mathbf{C}_{i,j}$ counts the links that cross between the halves $S_1$ and $S_2$ and the cut **C(X)** is 1/2 of the sum[11] of the $\mathbf{C}_{i,j}$ over $i,j=0..N-1$:[12]

$$\mathbf{C(X)} = \tfrac{1}{2}\sum_{i,j=0}^{N-1} \mathbf{C}_{i,j} = \tfrac{1}{2}\sum_{i,j=0}^{N-1} \tfrac{1}{2}(1 - x_i x_j) \mathbf{A}_{i,j} =$$

$$= \tfrac{nm}{4} - \tfrac{1}{4}\sum_{i,j=0}^{N-1} x_i x_j \mathbf{A}_{i,j} = \tfrac{n}{4}\left(m - \tfrac{\langle \mathbf{X}|\mathbf{A}|\mathbf{X}\rangle}{\langle \mathbf{X}|\mathbf{X}\rangle}\right) \quad (3.1)$$

Since bisection **B** is the minimum cut **C(X)** over all $\mathbf{X}\in\mathbf{E}$ (**E** is the set of all equipartitions), then via eq. (3.1):

$$\mathbf{B} = \min_{\mathbf{X}\in\mathbf{E}}\left\{\tfrac{N}{4}\left(m - \tfrac{\langle \mathbf{X}|\mathbf{A}|\mathbf{X}\rangle}{\langle \mathbf{X}|\mathbf{X}\rangle}\right)\right\} \equiv \tfrac{N}{4}(m - \mathbf{M}_\mathbf{E}) \quad (3.2)$$

$$\text{where:} \quad \mathbf{M}_\mathbf{E} \equiv \max_{\mathbf{X}\in\mathbf{E}}\left\{\tfrac{\langle \mathbf{X}|\mathbf{A}|\mathbf{X}\rangle}{\langle \mathbf{X}|\mathbf{X}\rangle}\right\} \quad (3.3)$$

Except for the constraint $\mathbf{X}\in\mathbf{E}$ (instead of $\mathbf{X}\in\mathbb{V}_N\supset\mathbf{E}$), expression (3.3) for $\mathbf{M}_\mathbf{E}$ looks the same as the Rayleigh quotient $\mathbf{M}_V$ in eq. (2.22), which is solved as $\mathbf{M}_V=\lambda_{max}$ by some eigenvectors $\mathbf{X}_0$ of **A**.

For *regular graphs* (fixed $m$), $\lambda_{max}$ of **A** is solved trivially by the eigenvector $\langle \mathbf{1}|\equiv(1\ 1\ldots 1)$, yielding via eq. (2.25): $\lambda_{max}= m$. Since $\langle \mathbf{1}| \notin \mathbf{E}$ this solution *does not apply* to (3.2). Hence, we will remove this eigenvector via decomposition: $\mathbb{V}_N=\mathbb{V}_1\oplus\mathbb{V}_{N-1}$, where $\mathbb{V}_1$ is subspace of $\mathbb{V}_N$ spanned by $\langle \mathbf{1}|$ and $\mathbb{V}_{N-1}$ is its orthogonal complement. Since $\langle \mathbf{1}|\mathbf{X}\rangle=0$ for all $\mathbf{X}\in\mathbf{E}$, all $\mathbf{X}\in\mathbf{E}$ are vectors of $\mathbb{V}_{N-1}$ i.e. $\mathbf{E}\subset\mathbb{V}_{N-1}$. Hence the max{} in (3.3) is constrained case of the general max{} in eq. (2.22) for $\mathbb{V}_{N-1}$. This implies via (**M₄**) that $\lambda_{N-1}\geq \mathbf{M}_\mathbf{E}$,[13] which via eq. (3.2) yields:

$$\mathbf{B} \geq \tfrac{N}{4}\cdot(m - \lambda_{N-1}) \quad (3.4)$$

The equality in eq. (3.4) holds iff the eigenvector $\mathbf{X}_0$ for $\lambda_{N-1}$ is an equipartition i.e. $\mathbf{X}_0\in\mathbf{E}$. A natural next step is to find graphs for which the equality condition holds and which allow for efficient eigen-decomposition algorithms.

### 3.2 Bisection for Cube-like Graphs

Regular $d$-cube ($d$ dim. hypercube, $Q_d$) is $Cay(\mathbf{Z}_2^d,\mathbf{S}_d)$ with bisection $\mathbf{B}=N/2$. Defining ***normalized bisection*** as $\mathbf{b}\equiv \mathbf{B}/(N/2)$, for $d$-cube $\mathbf{b}=1$. Folded $d$-cube $FQ_d$, which is **B** and distance optimal $Cay(\mathbf{Z}_2^d,\mathbf{S}_{d+1})$, has $\mathbf{b}=2$. The remarkable effectiveness of the $FQ_d$ augmentation of $Q_d$, which doubles **B** and halves diameter **D** of $Q_d$ while adding only one link per node ($m: d \to d+1$), motivated the exploration of the general[14] $Q_d$ extension of this type:

---

[8] We measure cuts and bisections in link units.
[9] The link changes are within the given class of topologies.
[10] The $N$ nodes are labeled as integers 0,1,… $N$-1.
[11] Factor 1/2 is due to the fact that $\mathbf{C}_{i,j}=\mathbf{C}_{j,i}$ count the same link, hence the sum over all $i$ and $j$ counts each $i,j$ link twice.
[12] Via $\langle \mathbf{X}|\mathbf{X}\rangle = \sum_{i=0}^{N-1} x_i x_i = N$ and (2.26) $\sum_{i,j=0}^{N-1}\mathbf{A}_{i,j} = N\cdot m$.
[13] Since $\mathbf{M}_\mathbf{E}$ is a max{} over **X** from a proper subset $\mathbf{E}\subset\mathbb{V}_{N-1}$, while $\lambda_{N-1}$ is max{} over all vectors from $\mathbb{V}_{N-1}$.
[14] For further generalizations, including $Cay(\mathbf{Z}_q^d,\mathbf{S}_m)$ extending hyper-torus of length $q$ and dimension $d$, see [25].

$$XQ_{d,m} \equiv Cay(\mathbf{Z}_2^d, \mathbf{S}_m) \quad for \; d \leq m < N = 2^d \quad (3.10)$$

The **B**-optimal $XQ_{d,m}$ was unknown and there wasn't even a tractable algorithm for **B**. Solutions to both problems follow, starting with $O(N \cdot \log(N))$ exact algorithm for **B**.

$XQ_{d,m}$ has $N=2^d$ nodes denoted as $d$-bit integers $x \in [0..N)$. A node $x$ is connected to $m$ other nodes $y_s = x \wedge h_s$, s=1..$m$, where $m$ generators (hops) $h_s \in \mathbf{S}_m$ are also $d$-bit integers. Since node $x=0$ is connected to nodes $h_1, h_2,.. h_m$, the row 0 of adjacency matrix **A** has $m$ elements $\mathbf{A}(0,h_s) \equiv A_{0,h_s} = 1$ (for s:1..$m$) and the rest is 0. A general row $x$ has $m$ non-zero elements $\mathbf{A}(x, x \wedge h_s) = 1$. Denoting contributions of a single generator $h \in \mathbf{S}_m$ to **A** as $N \times N$ matrix $\mathbf{T}(h)$, **A** can be expressed more concisely using Iverson brackets as:

$$\mathbf{T}(h)_{i,j} \equiv [i \wedge j = h] \quad (3.11)$$

$$\mathbf{A} = \sum_{h \in \mathbf{S}_m} \mathbf{T}(h) = \sum_{s=1}^{m} \mathbf{T}(h_s) \quad (3.12)$$

Few useful properties of $\mathbf{T}(h)$ matrices are (via (3.11)):

$$\mathbf{T}(a)_{i,j} = \mathbf{T}(a)_{j,i} \quad (3.13)$$

$$\mathbf{T}(a)\mathbf{T}(b) = \mathbf{T}(a \wedge b) \quad (3.14)$$

$$\mathbf{T}(a)\mathbf{T}(b) = \mathbf{T}(b)\mathbf{T}(a) \quad (3.15)$$

$\mathbf{T}(h)$ are thus symmetric, mutually commuting matrices and are representation of $\mathbf{Z}_2^d$. From ($\mathbf{M}_2$) it follows that $\mathbf{T}(h)$ have a common, complete eigenbasis. We now show (via (2.5)) that the $N$ Walsh vectors $|\mathbf{U}_r\rangle$ are this common, complete eigenbasis for all matrices $\mathbf{T}(a)$, $a \in \mathbf{Z}_2^d$:

$$(\mathbf{T}(a)|\mathbf{U}_r\rangle)_i = \sum_{j=0}^{N-1} [i \wedge j = a] \, \mathbf{U}_r(j) = \mathbf{U}_r(i \wedge a) \quad (3.16)$$

$$\mathbf{U}_r(i \wedge a) = (-1)^{\sum_{\mu=0}^{d-1} r_\mu (i \wedge a)_\mu} = (-1)^{\sum_{\mu=0}^{d-1} (r_\mu a_\mu + r_\mu i_\mu)} =$$

$$= (-1)^{\sum_{\mu=0}^{d-1} r_\mu a_\mu} \cdot (-1)^{\sum_{\mu=0}^{d-1} r_\mu i_\mu} =$$

$$= \mathbf{U}_r(a)\mathbf{U}_r(i) = \mathbf{U}_r(a)(|\mathbf{U}_r\rangle)_i \quad (3.17)$$

Collecting the $N$ components $i$ on l.h.s. of (3.16) and r.h.s. of (3.17) and expressing them in vector form yields:

$$\mathbf{T}(a)|\mathbf{U}_r\rangle = \mathbf{U}_r(a)|\mathbf{U}_r\rangle \quad (3.18)$$

Since matrix **A** commutes with all $\mathbf{T}(h)$ matrices, eqs. (3.12), (3.18) solve the eigenproblem of **A** as follows:

$$\mathbf{A}|\mathbf{U}_r\rangle = \left(\sum_{s=1}^{m} \mathbf{U}_r(h_s)\right) \cdot |\mathbf{U}_r\rangle \equiv \alpha_r |\mathbf{U}_r\rangle \quad (3.19)$$

$$where: \; \alpha_r \equiv \sum_{s=1}^{m} \mathbf{U}_r(h_s) \quad (3.20)$$

$$\alpha_0 = m \geq \alpha_r \quad for \; r > 0 \quad (3.21)$$

Thus $\alpha_0$ is the trivial (max) eigenvalue with eigenvector $\langle \mathbf{U}_0| = \langle \mathbf{1}|$, as in general regular graph case. The nontrivial $N$-1 eigenvalues $\alpha_r$ for $r>0$ have, via (3.19), eigenvectors $\mathbf{U}_r$ which via eq. (2.7) are also *equipartitions* $\mathbf{U}_r \in \mathbf{E}$. Hence eq. (3.4) applies with equality, solving for **B**:

$$\mathbf{B} = \frac{N}{4}\left(m - \max_{r>0} \alpha_r\right) = \frac{N}{4}\left(m - \max_{r>0} \left\{\sum_{s=1}^{m} \mathbf{U}_r(h_s)\right\}\right) \quad (3.22)$$

For programming and optimization of **B**, the binary form $\mathbf{W}_r$ of $\mathbf{U}_r$ is more convenient. We translate **B** algorithm of eq. (3.22) into the binary form using eq. (2.8):

$$\mathbf{B} = \frac{N}{4}\left(m - \max_{r>0}\left\{\sum_{s=1}^{m}(1 - 2 \cdot \mathbf{W}_r(h_s))\right\}\right) = \frac{N}{2}\min_{r>0}\left\{\sum_{s=1}^{m}\mathbf{W}_r(h_s)\right\} \Rightarrow$$

$$\mathbf{b} \equiv \frac{\mathbf{B}}{N/2} = \min_{r>0}\left\{\sum_{s=1}^{m}\mathbb{P}(r \& h_s)\right\} \quad (3.23)$$

From eqs. (3.23) and (3.2) we can interpret the sum being minimized in (3.23) as the $cut(\mathbf{W}_r)$ (in units $N/2$) of the binary partition vector $\mathbf{X} = \mathbf{W}_r$ (1s and 0s of $\mathbf{W}_r$):

$$C_r \equiv cut(\mathbf{W}_r) = \sum_{s=1}^{m}\mathbf{W}_r(h_s) = \sum_{s=1}^{m}\mathbb{P}(r \& h_s) \quad (3.24)$$

Hence algorithm (3.23) replaces the cut evaluations over $O(2^N)$ partition vectors $\mathbf{X} \in \mathbf{E}$ with cut $C_r$ evaluations over only $N$-1 partition vectors corresponding to Walsh function patterns. Besides the major savings in number of partitions checked, (3.23) also reduces the work for each cut $C_r$ itself to addition of $m$ terms vs. the general algorithm in eq. (3.1) which adds $N \cdot m$ terms.

A direct and simple C implementation of (3.23) is shown below. The inner loop in (3.25) executes $N \cdot m$ times. This can be further optimized via Fast Walsh Transform to run in $O(N \cdot \log(N))$ time (cf. [27] p. 24).

```
int Bisection(int N,int *hops,int m)           (3.25)
{
int cut,b,i,r;
  for(b=N,r=1; r<N; ++r)      // Check all Wr()
    {
    for(cut=i=0; i<m; ++i)    // calc cut(Wr)
      cut+=Parity(r&hops[i]); // via eq. (3.24)
    if (cut<b) b=cut;         // keep min cut b
    }
  return b; // Return bisection in units N/2
}
// Parity of 32-bit integer x, cf. [11]
inline int Parity(unsigned int x)              (3.26)
{
  x^=x>>16, x^=x>>8, x^=x>>4, x^=x>>2;
  return (x^(x>>1))&1;
}
```

### 3.3 Optimizing Bisection

Direct optimization of **B** requires evaluating (3.23) for all sets $\mathbf{S}_m = \{h_1, h_2, \ldots h_m\}$ of $m$ hops to find the set with maximum **B**. With $O(N^{m-d})$ such $\mathbf{S}_m$ sets[15], the overall complexity is $O(N^{m-d+1}\log(N))$ which is polynomial in $N$, hence tractable in principle. In practice, the polynomial degree ($m$-$d$+1) limits the sizes $N$ and link densities $m$ for which such brute force approach is usable. Much faster, greedy algorithms which iteratively replace 1 or 2 hops from $S_m$ at a time, resulting in O($N^2$) or O($N^3$) complexity, yielded fairly good solutions during the initial explorations. But that approach left unclear how close these solutions were to the exact optima and when could the search be terminated.

---

[15] The first $d$ hops can be kept fixed as hypercube basis without a loss of generality, cf. ($EC_2$).

Entirely different way for optimizing **B** emerges from closer examination of the expression (3.24) for the cut $C_r$ which is illustrated below for XQ$_{4,5}$ with $d$=4, $m$=5 hops, and cut $C_r$ for $r$=0xB=1011. The hop list $S_m$ is shown and interpreted as a bit matrix of dimensions $m{\times}d$.

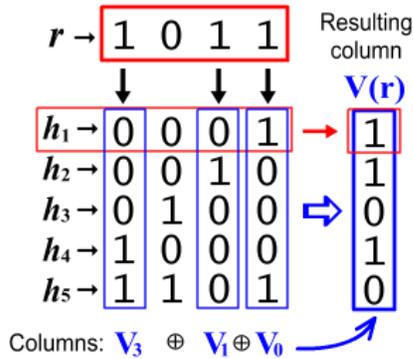

**Figure 3-2: Bit columns action of Walsh function $W_r$**

The results of each term in (3.24), $\mathbb{P}(r\,\&\,h_s)$, are shown in the column **V**($r$). For a single row, the expression $\mathbb{P}(r\&h_s)$ computes linear combination $\sum_{\mu=0}^{d-1} r_\mu \cdot (h_s)_\mu$ in GF(2) to get a bit for that row in column **V**($r$). Hence, the full column vector $|\mathbf{V}(r)\rangle$ is a linear combination in GF(2)$^m$ of the bit columns $|\mathbf{V}_\mu\rangle \in \mathbb{V}_m$ of the hop list $S_m$, and the sum in (3.24) is the "Hamming weight of $|\mathbf{V}(r)\rangle$" $\equiv \Delta\mathbf{V}(r)$:

$$C_r = \sum_{s=1}^{m} \mathbb{P}(r\&h_s) = \Delta\left(\sum_{\mu=0}^{d-1} r_\mu |\mathbf{V}_\mu\rangle\right) = \Delta\mathbf{V}(r) \quad (3.30)$$

Bisection **b** from eq. (3.22) is in this formulation given as:

$$\mathbf{b} = \min_{r>0}\{\Delta\mathbf{V}(r)\} \quad (3.31)$$

The set of vectors **V**($r$) in (3.31) is $|\mathbf{V}(r)\rangle = \sum_{\mu=0}^{d-1} r_\mu |\mathbf{V}_\mu\rangle$ i.e. set $\mathbb{V}_d \equiv \{\mathbf{V}(r): 0 \leq r < N\}$ is a $d$-dimensional subspace $\mathbb{V}_d \subset \mathbb{V}_m$. The **B** optimization is then a problem of finding a subspace $\mathbb{V}_d \subset \mathbb{V}_m$ which maximizes **b** from (3.31):

$$\mathbf{b}_{opt} = \max_{\mathbb{V}_d \subset \mathbb{V}_m}\left\{\min_{\mathbf{V}\neq 0}\{\Delta\mathbf{V}: \mathbf{V} \in \mathbb{V}_d\}\right\} \quad (3.32)$$

Except for the labels, $\mathbf{b}_{opt}$ in (3.32) is identical to the problem of $\Delta_{opt}$ in (2.36) i.e. the two problems are mathematically one and the same.

Hence, the translation recipe for converting between [$\_n,\_k$] codes[16] over GF(2) and Cayley graphs $Cay(\mathbf{Z}_2^d,S_m)$[17] is as follows:

**Table 3-1. Equivalence EC Codes ↔ Networks**

| [$\_n,\_k,\Delta$] code | $\_n$ | $\_k$ | $\Delta$ | $\mathbf{G}^T$ | $|g_i\rangle$ | **X** | **Y(X)** | $\Delta\mathbf{Y}$ |
|---|---|---|---|---|---|---|---|---|
| $Cay(\mathbf{Z}_2^d,S_m)$ | $m$ | $d$ | b | $S_m$ | $|\mathbf{V}_\mu\rangle$ | $r$ | **V**($r$) | $C_r$ |

Examples: repetition code ↔ trunking (LAG), parity bit code ↔ folded hypercube, Hadamard code ↔ fully connected graph, Reed-Muller code RM(1,$d$-1) ↔ Turán graph T($N$,2) or complete bipartite graph K$_{N/2,N/2}$.

### 3.3.1 Construction Recipe: EC Codes → Networks
To construct optimal bisection networks from optimal $\Delta$ EC codes one would start with network specification such as $N$=2$^d$ switches (which yields $d$=log($N$)) and $m$ topological (switch-switch) ports/switch.

C1. For given network parameters $d$ and $m$ find[18] the best available [$\_n$=$m,\_k$=$d$] code over GF(2) and its generator matrix **G** of size $d{\times}m$ ($d$ rows, $m$ columns). The result is [$\_n,\_k,\Delta$] code which has the largest min. codeword distance $\Delta$ for given $\_n$ and $\_k$.

C2. Transpose **G** (or rotate it 90°) to get $m{\times}d$ matrix $S_m$ and read the $m$ hops $h_s$, each as $d$ binary digits per row of $S_m$ (see Figure 3-2)[19].

C3. Label $N$ network nodes (switches) as 0,1,…$N$-1 and for node $x$ compute the $m$ nodes $y_1, y_2, …\ y_m$ linked to $x$ using: $y_s = x{\wedge}h_s$ for $s$=1,2…$m$.

C4. Network bisection (in link units) is **B**=$\Delta{\cdot}N/2$ which provides $\Delta$ non-oversubscribed ports on each switch.

## 3.4 Long Hop Networks
The networks constructed from the optimal codes via the above recipe were named *Long Hop* networks (**LH**). The current LH solutions data base contains 3364 solutions extending to $N$=2$^{20}$ switches and to $m$ = 256 topological ports/switch, yielding networks with up to $P$ = 117$\cdot$10$^6$ non-oversubscribed ports using radix $R$=384 switches.

We next compare LH with 5 popular or proposed network topologies (for formulas used and spreadsheets cf. [28] [29]), some contending for the best performing networks (cf. [1], [15], [30] [31], [32], [33], [34]). All networks were set to use the same radix switches and generate as efficiently as possible the same number of non-oversubscribed ports (i.e. to have the same bisection). We then compare the total numbers of switches (as

---
[16] To avoid mix-up of notations, the ECC symbols $n$, $k$ are denoted as $\_n$ and $\_k$ in this section.
[17] For generalization to $Cay(\mathbf{Z}_q^d,S_m)$ (generalized hyper-torus) from non-binary EC codes over GF($q$), $q$ >2 see [25].
[18] E.g. via repositories [18], [19] and MAGMA package
[19] If the $S_m$ doesn't have hypercube basis $h_s$=2$^s$ it can be diagonalized via linear combinations of columns, cf. (EC$_2$).

Ports/Switch ratio, higher is better) and topological cables (as Cables/Port ratio, lower is better) needed for the task. Since networks had different 'natural' configurations that don't yield exactly the same number of ports, in the charts the alternative networks generate their 'natural' optimal sizes (ports and switches), then we interpolate between the nearest higher/lower LH configurations for the target number of ports. In Table 3-2 we reverse the roles and use specific LH topology, then interpolate between nearest optimal configurations of the alternatives to obtain the same number of non-oversubscribed ports (reaching the same conclusions).

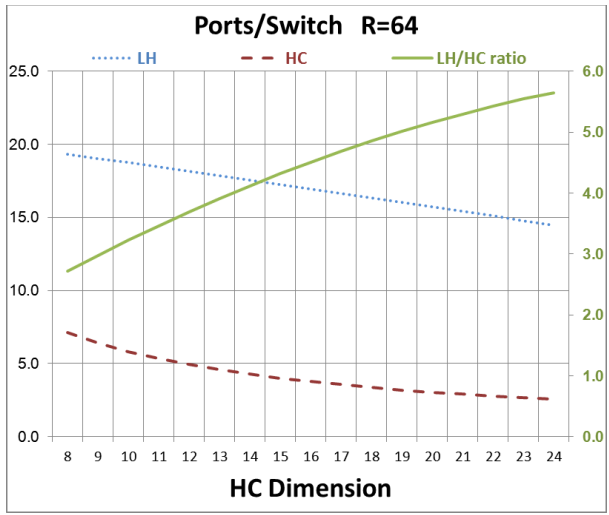

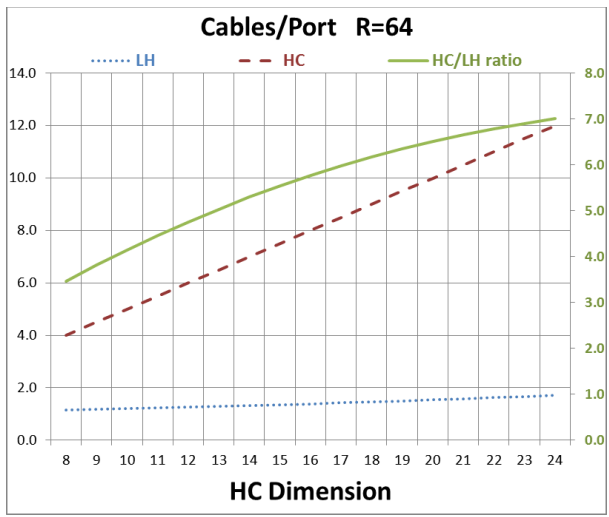

Figure 3-3: LH vs. Hypercube

For each chart, the left scale shows the value of quantity compared while the right scale shows the LH advantage ratio. E.g. Figure 3-3 shows that for hypercube with $N=2^8-2^{24}$ switches, LH providing the same number of non-oversubscribed ports yields 2.7-5.7 times more ports/switch (or using 2.7-5.7 times fewer switches) while using 3.5-7 times fewer cables than hypercube.

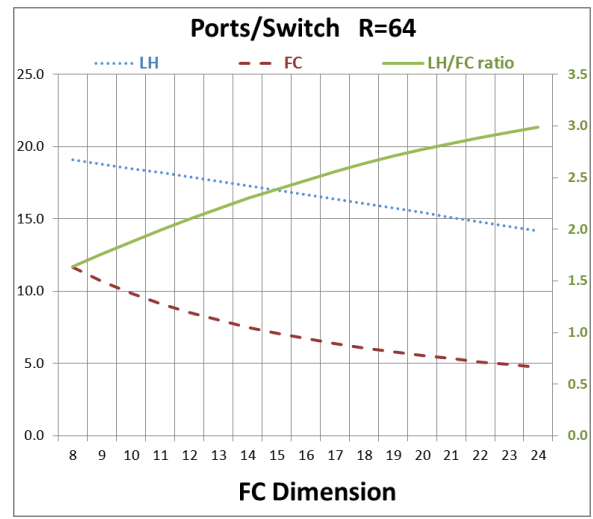

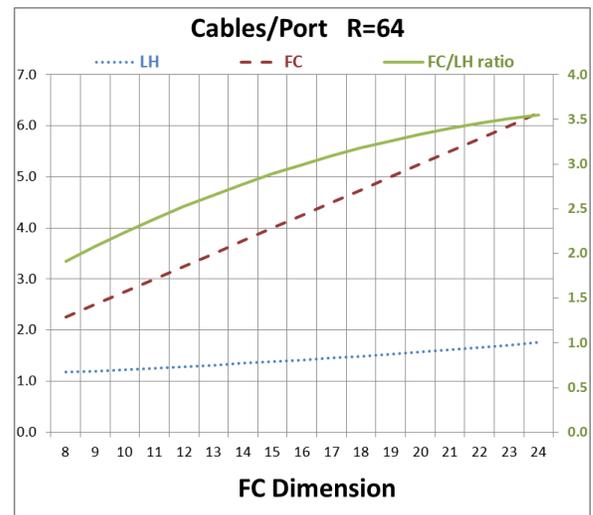

Figure 3-4: LH vs. Folded Cube

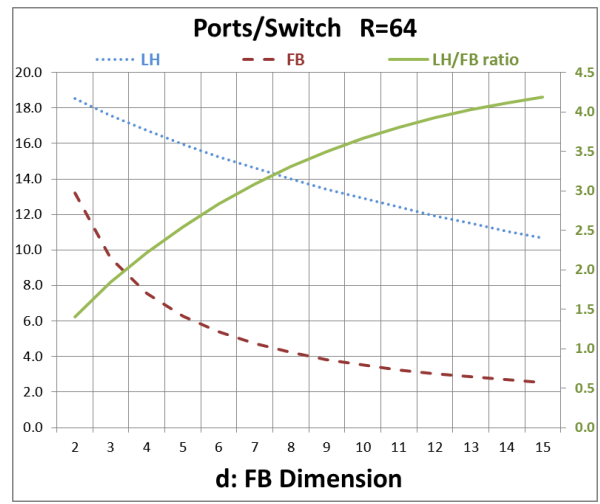

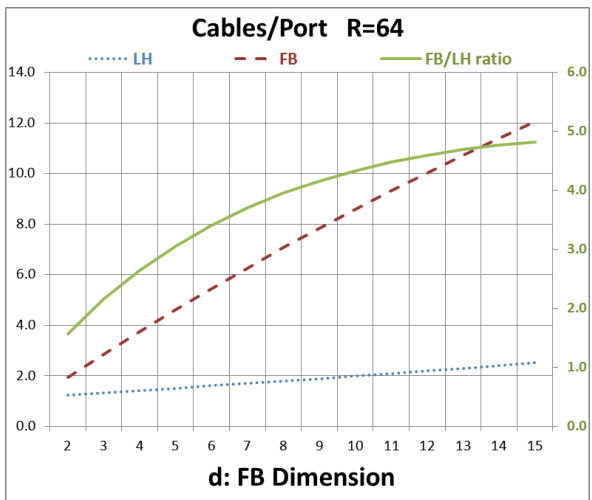

Figure 3-5: LH vs. Flattened Butterfly

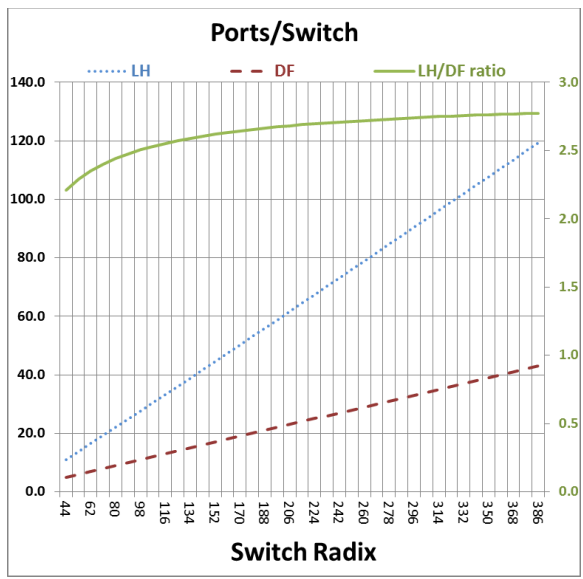

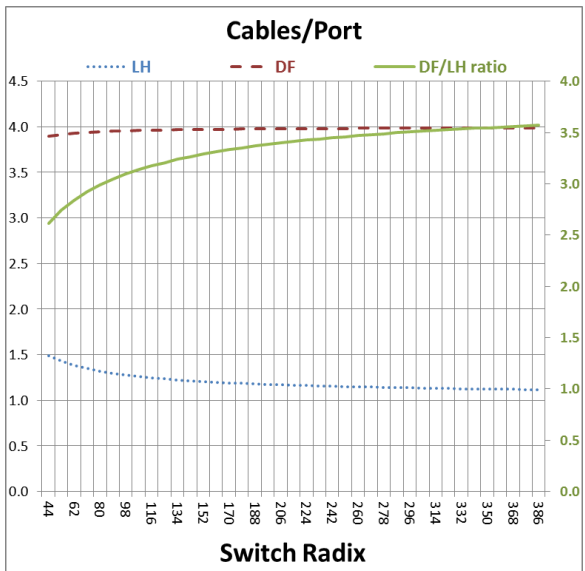

Figure 3-6: LH vs. Dragonfly

In comparison with Dragonfly (DF), we vary switch radix for both networks since optimal non-oversubscribed DF (which is max DF) lacks any other free parameters for changing the network size but switch radix.

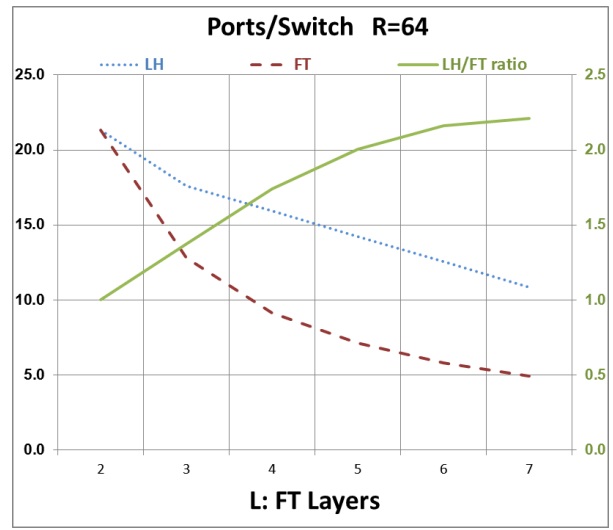

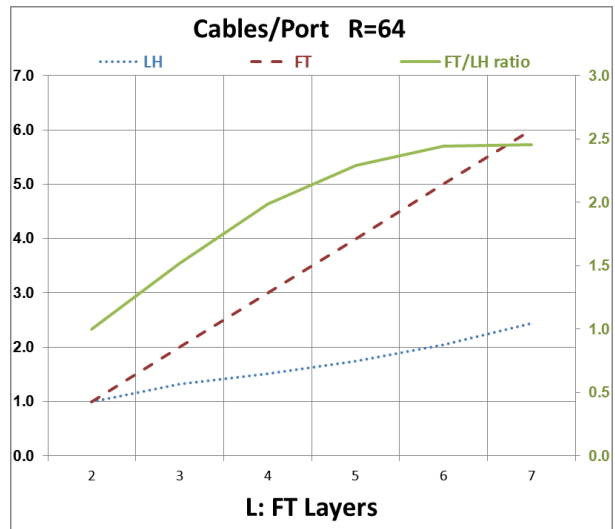

Figure 3-7: LH vs. Fat Tree

The Fat Tree (FT) comparisons use non-trunked (max) FT for each number of FT levels, which is the most efficient FT (LH advantage ratios are larger for trunked FT) . The two level FT (FT-2), being a complete bipartite graph, has optimal bisection, hence it yields the same figures for Ports/Switch = $R$/3 and Cables/Port = 1 as LH. For network sizes beyond the reach of FT-2 (i.e. when number of switches exceeds 1.5·$R$), the bisection of FT-L, L>2, is not optimal any longer and the LH advantage ratios increase with the number of FT levels.

Since the charts set both networks to common bisection (bottleneck), the networks are normalized to the same *worst case traffic* which misses the *major weakness* of Fat Tree for *random or benign* traffic (a far more probable traffic than the worst case traffic) – random traffic throughput of FT is the same as its worst case throughput, while all other networks compared have 1.5-2 times larger capacity for random or benign traffic than for the worst case traffic. This FT problem is shown in Figure 3-8 (cf. [33], p.6) in chart (a) where in contrast to hypercube and Flattened butterfly, FT saturates at 50% of network capacity for random traffic. Hence, if networks were normalized to the same random traffic performance, the LH advantage ratios vs. FT, shown on the right scales in Figure 3-7, would increase by a factor 1.5-2×.

A more detailed comparison is shown in Table 3-2 (output from TCALC program, [28]), where a specific LH network, yielding ~131K ports is compared to the 5 alternatives. Shaded columns normalize costs for cabling and switches so LH is 100. The column "Cost Gb/s" normalizes all networks to the same random/benign traffic throughput. As result, the FT comes behind not just LH but also behind FB and DF. That is how papers [33], [34] have compared these three networks showing similar performance advantages of FB and DF vs. FT.

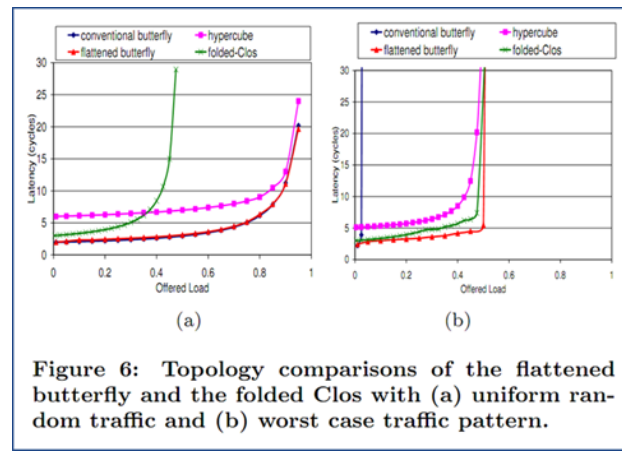

Figure 6: Topology comparisons of the flattened butterfly and the folded Clos with (a) uniform random traffic and (b) worst case traffic pattern.

**Figure 3-8: FT Overload on random traffic**

**Table 3-2: Comparison of a specific LH network with alternative topologies**

| | | | | | | | | |
|---|---|---|---|---|---|---|---|---|
| NETWORKS COMPARED WITH THE LONG HOP (LH) NETWORK ||||||||||
| Fat Tree (FT): FT Levels L=4   Q=2.519842 (trunking factor) <br> Flattened Butterfly (FB): FB(k:17.428, n:4.365, c:8.714) <br> Dragonfly (DF): DF(p:7.24, a:37.85, h:12.62, g:478.44)   Q=1.148 <br> Folded Hypercube (FC): FC(dimension 14.096)  Q=3.744 <br> Hypercube (HC): HC(dimension 15.000)  Q=4.000 ||||||||||
| TARGET: **Ports** P=**131,072**,   Switch **radix** R=**64**,   Oversubscription **ovs=1** ||||||||||
| ## | #Switches | Ports/Sw. | Switches | Cost Gb/s | Cables/Pt | Cabling | Max | Avg Hops | Latency |
| LH | 8192 | 16.000 | 100 | 100 | 1.500 | 100 | 4 | 2.915039 | 100 |
| FT | 14336 | 9.143 | 175 | 358 | 3.000 | 200 | 6 | 5.968750 | 205 |
| FB | 15042 | 8.714 | 184 | 238 | 3.172 | 211 | 4 | 3.777778 | 130 |
| DF | 18107 | 7.239 | 221 | 221 | 3.921 | 261 | 3 | 2.916464 | 100 |
| FC | 17506 | 7.487 | 214 | 447 | 3.774 | 252 | 8 | 6.100012 | 209 |
| HC | 32768 | 4.000 | 400 | 1029 | 7.500 | 500 | 15 | 7.500000 | 257 |

## 4. FLEXIBLE RADIX SWITCH

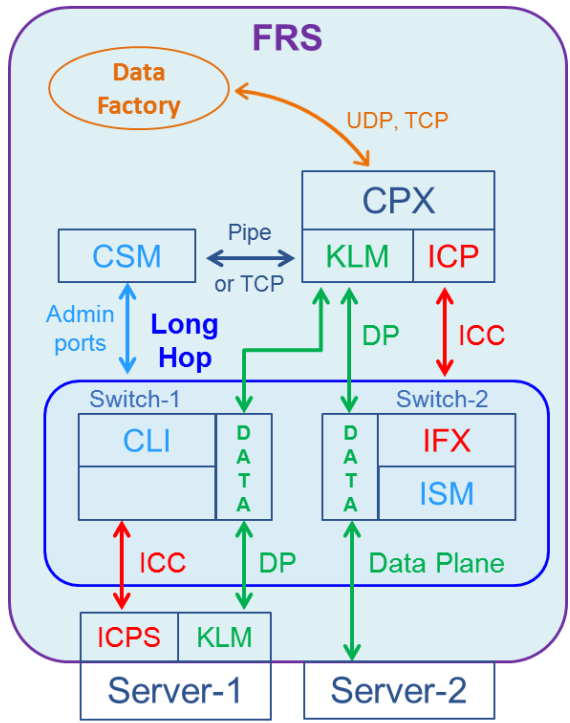

**Figure 4-1: FRS Architecture**

For networks which support general topology, such as InfiniBand (IB) or HPC systems, LH deployment should be simple, requiring at most, as an optimization, the integration of LH library for routing and forwarding computations into the IB Subnet Manager. Similarly a pure Layer 3 (L3) deployment as a replacement for Fat Tree e.g. under OSPF or BGP for management of L3 topology, would be unproblematic, although for large Data Centers that approach would not extract most of the gains available to FRS such as those in Figure 1-2.

The principal difficulty in implementing general topology on Ethernet is in gaining control over its flood based forwarding and the ARP broadcasts and replacing them with deterministic single path alternatives. FRS combines methods most similar to those of Portland [6], Triton [2] and NetLord [7]. For greater deployment flexibility, FRS implements two modes of operation regarding forwarding control: '**CLI mode**' (command line interface via switch admin ports[20]) + server shim, or '**Switch mode**' with FRS components running on switches as 'switch agents'. The latter mode doesn't require server components although it still uses them whenever possible for added flexibility. The control paths for the two modes are indicated in Figure 4-1 via suffix -1 or -2 on switch/server labels.

### 4.1 FRS Components

The top level central controller for FRS is **CPX** program (control plane executive) which starts and controls its network facing components, **ICP** (integrated control plane) and **KLM** (Linux kernel loadable module, shim between L3 & L2) and interfaces them to the data base and management software, Data Factory[21].

The main networking component is ICP (analogous to IB Subnet Manager) which controls its 'satellites' **ICPS** (on servers), **IFX** (on switches) and **KLM** (in kernel or in hypervisor).

Switch *hardware abstraction* is implemented by the two types of 'Switch Manager' (SM) modules, the central SM, **CSM**[22] which controls switches through admin ports via CLI and internal SM used by IFX, **ISM**, which interfaces to switch vendor's API for switch agents such as EOS on Arista switches.

**ICC** is ICP's *control channel* for messages with ICPS and IFX. Although depicted above as logically separate from Data Plane (DP), physically the ICC runs over the same DP as regular data. The KLM which communicates only with the user mode programs on the same computer, uses pipes. CPX communicates with ICP via pipes and with Data Factory via TCP and UDP. CPX controls KLM only indirectly via ICP.

All I/O within FRS components uses non-blocking descriptors and sockets via event driven poll/select mechanism, which provides a fast, light-weight context switching between multiple I/O channels without unnecessary thread or process switching overhead per event[23]. The priority queues running in the same event loops can handle tens of thousands of pending events with $O(1)$ dequeue time and $O(\log(n))$ add-event time.

### 4.2 Operational Elements

#### 4.2.1 Long Hop Paths

FRS uses non-minimal multipath routing and forwarding. Path computations on LH are almost as simple as those on hypercube (HC). Thanks to vertex symmetry, the paths and resulting forwarding tables need to be computed only from one node X=0 to all other nodes Y. The relative paths (hop sequences) from node X≠0 to Y are the same as the relative paths from 0 to X^Y.

The shortest paths, e.g. on 4-cube from X=0 to Y=1011 have 3 hops and there are 3!=6 paths (each bit=1 corresponds to a hop along 1 HC dimension and the 3 hops can be made in any order). In LH with ***m*** topological ports/switch, the paths from X=0 to some Y at distance

---

[20] FRS can also use OpenFlow in 'CLI mode' if available.

[21] Due to space constraints, we will focus on networking aspects.
[22] In the prototype, CSM runs on the CPX machine and talks to ICP via a pipe. On larger networks several CSM copies can run on separate servers using TCP for messages with ICP.
[23] One I/O thread is used per core available.

L=3 hops, can be partitioned into one or more *path sets* (number of paths sets depends on Y), with each path set operating like HC paths, except that the 3 ones (for L=3 hops) are within the *m*-bit string (each bit corresponds to one egress port) instead of in a *d*-bit strings for HC.

The construction of non-minimal paths is controlled by a parameter $Q \leq m$, which is the number of edge disjoint paths required between any two nodes. First, the shortest paths are computed via paths sets. If there are not Q such edge disjoint paths, the algorithm computes additional paths which are 1 hop longer than the shortest path, then if these don't reach the Q paths either, the 2 hop longer paths are included, etc.

The Q paths per destination Y (from X=0) are then encoded into a forwarding table using Q *aliases* per destination Y i.e. the aliases of Y are Q pairs (s,Y), where s=1..Q is a *path selector*. With maximum path diversity Q=*m*, for any given Y each value of s selects a different topological egress port (out of *m* available). Depending on deployment constraints, path selectors s use either a VLAN ID (thus using up Q VLAN IDs), or an alias field in the topological MAC address of the switch (if the switch supports multiple self-addresses).

For *N* switches and the Q aliases per destination, the number of switch-to-switch forwarding entries is $N \cdot Q$ (instead of *N*). We have found in simulations that at scales of practical interest $Q \cong 4\text{-}5$ will yield nearly all of the multipath gains in reducing congestion, hence the FIB burden from multi-pathing need not be excessive.

### 4.2.2 Basic Layer2+3 Forwarding
Regular L2 flooding on unknown Dst MAC address (MA), all broadcasts (such as ARP), STP and MAC learning are disabled on the switches. Taking advantage of the fact that DC is a managed network, only the known (to FRS) destination addresses are allowed into and are forwarded by FRS. The L2 static tables (FIBs) are programmed to forward from any switch only up to egress switch of the destination server, while the last hop to the server is done via L3 forwarding on Dst IP via the IP table (this method is used in [7]). Hence, the load on IP tables is not excessive since each egress switch only needs to know the IPs of the attached servers (including any VMs). The load on the L2 FIBs is reduced, compared to having to forward to all MAs in the network, by the switch fanout factor (typically 20-40 server ports/switch). If the L2 FIBs suffice for the network size[24], no further topological addressing (beyond the two levels above) is introduced.

### 4.2.3 ARP responses and Path Control
When the server KLMs are available, the ARPs from servers are disabled. The KLM intercepts all outgoing packets (to FRS interfaces) between L3 and L2 modules, right after the L3 headers were created. Based on Dst IP, KLM selects the correct Dst MA (of the egress switch for Dst IP and given Q), prepends the L2 header and sends the completed frame to the NIC driver, bypassing the default L2 processing (rendering ARP unnecessary).

In pure 'Switch mode' (without KLMs), the IFX+ISM on switches trap all ARP requests from attached servers, squelch them and respond with proper Dst MA as in KLM method above. Gratuitous ARPs are sent to servers for any updates of ARP tables.

In either mode, whenever multipath parameter Q>1, the new flows are spread out among the Q available paths.

### 4.2.4 Third level of topological addressing
For larger networks or larger multipath value Q, when the capacity of L2 FIBs is insufficient, a *third level* of topological addressing is added as '*cluster*' and '*cell*' (within cluster) address levels[25], [9]. When forwarding, on the 'cluster' field mismatch with current switch, the next hop is forwarded on the 'cluster' field, and on the matching (final) 'cluster' the hop is forwarded on the 'cell' field. This approach reduces the number of forwarding entries from $N$ (# of switches) to $2\sqrt{N}$. FRS implementation uses one of two mechanisms, depending on deployment constraints and resources:

a) The topological MA of the switches is split into 'cluster' and 'cell' fields forwarded via L2 TCAM.
b) The network is split into 'clusters' which are private FRS L3 subnets[26], each an L2 domain, while 'cells' are MAs within the domain. The forwarding at L3 to other 'clusters' is done via L3 TCAMs (LPM tables)[27], and at L2 to other 'cells' within the same cluster via L2 FIBs. In this mode the L3 ECMP is used to augment the L2 alias based multi-pathing, reducing thus Q value and the load on L2 FIBs.

### 4.2.5 Topology Management
The topology discovery is coordinated by ICP upon receiving network configuration messages from CPX. The full, live network model is maintained only by ICP, while servers or switches know only their nearest neighbors.

In CLI mode, CSM obtains LLDP neighborhood records from each switch and ICP uses this info to construct the LH topology (assign LH node IDs and create node records). Changes to topology are detected by CSM via SNMP traps and are updated incrementally by ICP. In Switch mode, ICP runs a much faster discovery and topology change detection protocol jointly with IFX modules on switches (which use modified LLDP with EtherType 0x99AA) and ICPS modules on servers (these are optional in Switch mode).

---

[24] Broadcom Trident has L2 FIBs with 128K entries.

[25] Optimal clustering of LH is constructed via recursive splits along bisecting cuts which are computed via function (3.25).
[26] These private FRS subnets are invisible to servers, see 4.2.6.
[27] This method allows FRS to take full advantage of powerful L3 switching features available in recent fabrics.

After constructing topology, ICP computes the ICC *distribution tree* (allowing each server or switch to send/receive to/from ICP). The forwarding tables for this tree[28] are loaded into the switches[29] and if server components are used (ICPS & KLM), the ICC broadcast is sent to all servers, to let them identify themselves and join the network. Also loaded are general L2 static (and optionally L2 TCAM) tables for forwarding from any to any switch. After obtaining IP addresses from servers, ICP updates the egress IP tables for the discovered servers. These tables are also updated when servers leave or enter the network. Failures of the topological links or switches, are similarly updated in the L2 and L3 tables. The notifications of topology changes or IP movements are sent via ICC to servers and/or switches.

*4.2.6 Private FRS IP space*

Several mechanisms above rely on L3 switching features which introduces topological constraints on IPs. In order separate the FRS topological IPs from LAN IPs used by servers and applications (retaining thus the full mobility and agility of LAN IPs provided by the flat L2, [9]), FRS uses a NAT-like *IP rewrites*[30] which keeps its IP space invisible to servers and applications (in this mode, border routers and load balancers use FRS IPs for their LAN addresses). On outbound packets, the KLM overwrites Dst IP (and updates L3 header checksum) with the corresponding topological FRS IP and the receiver replaces it with the LAN IP bound to that L3 flow[31].

In this way FRS virtualizes global LAN IP space via a more economical IP rewrite instead of encapsulation with additional L2 and L3 headers (such as the one used in NetLord, [7]). The method does not virtualize network for each tenant separately, which was an objective in [7].

The FRS IP space is also useful in situations where the access routers were eliminated by FRS along with their ARP tables for the LAN, Figure 1-2. If the border router lacks capacity to handle the large ARP tables for the entire LAN, the topological FRS IPs are used together with method 4.2.4-b to provide full LAN routing without burdening the border router with IPs of all servers.

## 5. ACKNOWLEDGMENTS
The author thanks entire Infinetics team, especially Chris Williams, Harry Quackenboss, Lionel Pelamourgues, Leigh Turner and John Day for stimulating discussions and analysis of Data Center and HPC issues, Chris Williams for simulator experiments, Mike Ford for details on Data Center switches and implementation of switch controllers, John Zavgren and Reed Lewis for Linux kernel and hypervisor code collaboration, Scott Benson for his support of the network topology research.

---

[28] These are much smaller tables than general all-to-all tables. The dummy IPs used for egress L3 hop to server ports are taken from a separate subnet within private FRS IP space.

[29] In switch mode without admin network, our hop by hop custom LLDP is used to distribute table entries to switches.

[30] Present implementation of FRS IPs requires server KLMs. NAT capable switches may be used for this in the future.

[31] These IP bindings operate similarly to NAT on routers.